\documentclass[preprint,12pt]{elsarticle}

\usepackage{graphicx}
\usepackage{amssymb}

\usepackage{lineno}

\usepackage{amsmath}
\usepackage[safe]{tipa}
\usepackage{hyperref}

\journal{Astronomische Nachrichten}

\begin{document}

\begin{frontmatter}


\title{The Deep Underground Neutrino Experiment -- DUNE: the precision
era of neutrino physics}


\author[label1,label2]{Ernesto Kemp}
\ead{kemp@ifi.unicamp.br}
\author{for the DUNE collaboration}
\address[label1]{``Gleb Wataghin" Institute of Physics, University of Campinas -- UNICAMP, 13083-859, Campinas, SP, Brazil}
\address[label2]{Fermi National Accelerator Laboratory, Batavia, IL 60510-0500, USA}
\cortext[cor1]{To appear in the proceedings of the STARS 2017 conference:\\ \url{https://indico.cern.ch/event/542644/}}



\begin{abstract}
The last decade was remarkable for neutrino physics. In particular, the phenomenon of neutrino flavor oscillations has been firmly established by a series of independent measurements. All parameters of the neutrino mixing are now known and we have elements to plan a judicious exploration of new scenarios that are opened by these recent advances. With precise measurements we can test the 3-neutrino paradigm, neutrino mass hierarchy and CP asymmetry in the lepton sector. The future long-baseline experiments are considered to be a fundamental tool to deepen our knowledge of electroweak interactions. The Deep Underground Neutrino Experiment -- DUNE will detect a broad-band neutrino beam from Fermilab in an underground massive Liquid Argon Time-Projection Chamber at an L/E of about $10^3$ km / GeV to reach good sensitivity for CP-phase measurements and the determination of the mass hierarchy. The dimensions and the depth of the Far Detector also create an excellent opportunity to look for rare signals like proton decay to study violation of baryonic number, as well as supernova neutrino bursts, broadening the scope of the experiment to astrophysics and associated impacts in cosmology. In this presentation, we will discuss the physics motivations and the main experimental features of the DUNE project required to reach its scientific goals.
\end{abstract}


\end{frontmatter}


\section{The DUNE and LBNF projects}
\label{S:1}

Although the Standard Model of particle physics presents a remarkably accurate description of the elementary particles and their interactions, it is known that the current model is incomplete and that a more fundamental underlying theory must exist. Results from the last decades, that the three known types of neutrinos have nonzero mass, mix with one another and oscillate between generations, implies physics beyond the Standard Model. The neutrino mass generation shows to be more complex than the Higgs mechanism embedded in the Glashow-Salam-Weinberg electroweak theory. The neutrino interactions have very small cross-sections, for this reason, neutrinos can only be observed and studied via intense neutrino sources and large detectors. Neutrino experiments can be a straight way (at lower cost) to test the fundamentals of electroweak interactions. Additionally, a large detector placed underground, in a low-background environment can be used to study rare process, as the proton decay or supernovae neutrinos bursts, in which neutrinos seem to play a major role in the core-collapse mechanism and the subsequent ejection of the star's matter. If the collapsing star is massive enough, it can also end as a black hole.
The following list summarizes the above remarks and their related open questions in neutrino physics: 

\begin{itemize}

\item[--]
\textbf{Matter-antimatter asymmetry:} Immediately after the Big Bang, matter and antimatter were created equally, but now matter dominates. 

\item[--]
\textbf{Nature's fundamental underlying symmetries: }the patterns of
mixings and masses between the particles of the Standard Model is not understood.

\item[--]
\textbf{Grand Unified Theories (GUTs):} Experimental results 
suggest that the physical forces observed today were unified into one force at the birth of
the universe. GUTs predict that protons should decay, a process that has never been observed.

\item[--]
\textbf{Supernovae:} How do supernovae explode and what new physics will we learn from a neutrino burst?
\end{itemize}

To address all of these questions the worldwide neutrino physics community is developing a long-term program to measure unknown parameters of the Standard Model of particle physics and search for new phenomena. The physics program will be carried out as an international, leading-edge, dual-site experiment for neutrino science and proton decay studies, which is known as the Deep Underground Neutrino Experiment (DUNE).  The infrastructure for the beam and the experimental sites constitutes the Long-Baseline Neutrino Facility (LBNF). Together, DUNE and LBNF represent a very ambitious program in neutrino and elementary particle science, not to mention the impacts of the expected outcomes in astrophysics and cosmology. 
DUNE will comprise two experimental sites. The first at the Fermilab site, hosting the world’s highest-intensity neutrino beam and a high-precision near detector. The second at the Sanford Underground Research Facility (SURF) 1300 km away in Lead (SD), where a 40 kton liquid argon time-projection chamber (LArTPC) far detector will be installed underground ($\sim$ 1500 m deep). Fermilab will also provide all of the conventional and technical facilities necessary to support the beamline and detector systems.

\subsection{The Beam}

\begin{figure}
\includegraphics[width=\linewidth]{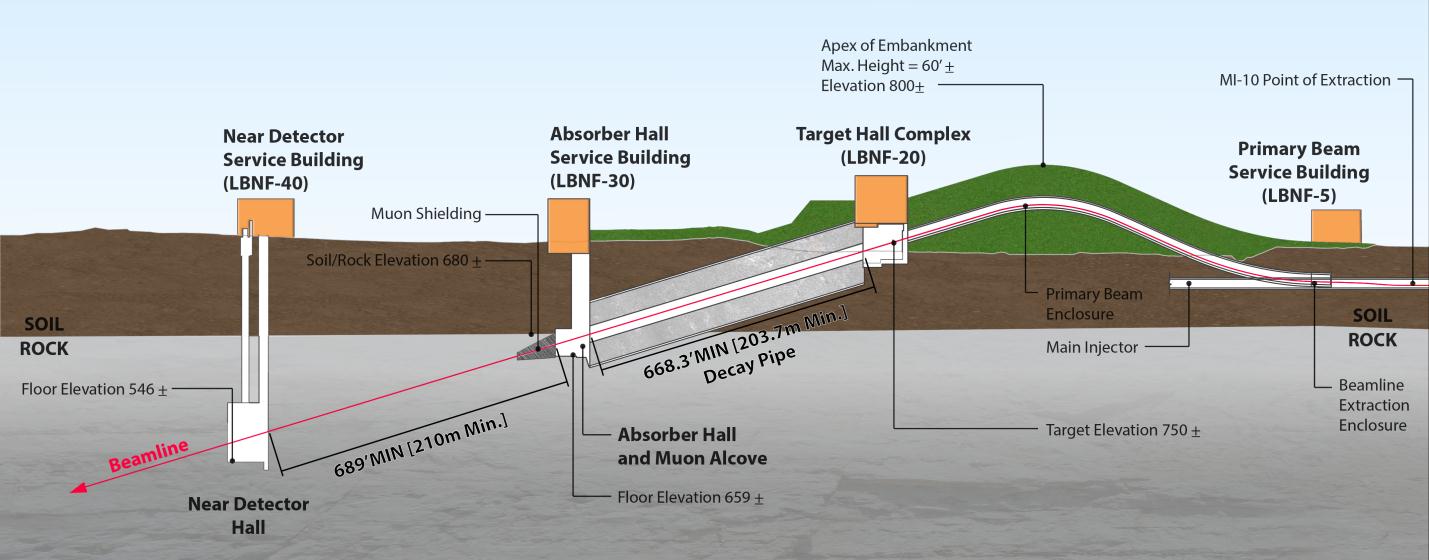}
\caption{Cross section of the beam line at Fermilab. From right to left: protons are extracted from the main injector and are pointed to SURF. The proton collisions at the target produce pions,  which decay into muons and neutrinos. Muons are filtered in the absorber and the remaining neutrino beam goes through the Near Detector and finally to the $\sim$ 1300 km long path through the Earth crust until the Far Detector.}
\label{beam}
\end{figure}

The LBNF beamline includes key features inherited from the successful NuMI beam line design for the MINOS and NOvA experiment (Ayres et al. 2004).  It exploits the same configuration of a target and horn system, with the spacing of the target and two horns tuned to obtain an intense neutrino flux at the first oscillation maximum and to extend as much as possible to the second as well. Profiting from the effective experience of the NuMI design the decay pipe is helium-filled, while the target chase is air-filled. The proton beam energy can be adjusted within the 60 and 120 GeV range, with the corresponding range of beam power from 1.0 to 1.2 MW. The ability to vary the proton beam energy is essential for optimizing the neutrino spectrum and to understand systematic effects in the beam production. An energy tunable neutrino beam also provides flexibility to allow the addressing of future questions in neutrino physics that may require a different neutrino energy spectrum. 
The reference design has values of 204 m length and 4 m diameter for the decay pipe, both matching well to the physics of
DUNE but studies to determine the optimal dimensions continue. The main elements of the beam line are shown in Figure \ref{beam}.

\subsection{The Far Detector}

\begin{figure}
\includegraphics[width=\linewidth]{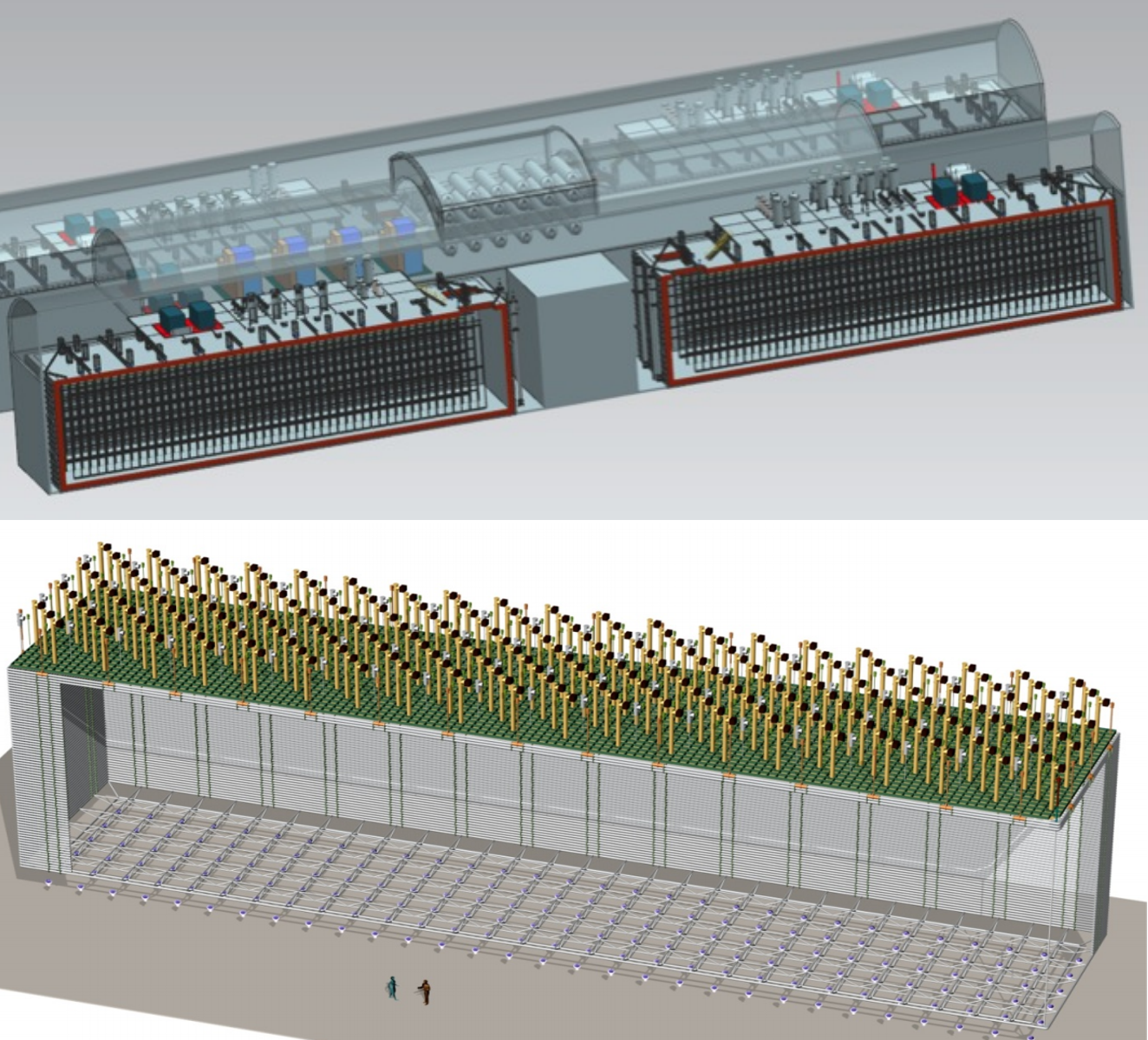}
\caption{3D models of two 10 kton detectors using the single-phase reference design (top) and the dual-phase alternative design (bottom) for the DUNE far detector to be located at 4850L.}
\label{fd-2detectors}
\end{figure}

The far detector (FD) will be composed of four similar modules, each one a liquid argon time-projection chamber (LArTPC). The LArTPC technology (Rubbia 1977; Acciarri et al. 2015a) provides excellent tracking and calorimetry performance, making it as an excellent choice for massive neutrino detectors such as the DUNE FD.
Moreover, the LArTPC ability for precise reconstruction of the kinematical properties of particles increases the correct identification and measurements of neutrino events over a wide range of energies. The full imaging of events will allow study of neutrino interactions and other rare events with an unprecedented resolution. The huge mass will grant the collection of a vast number of events, with sufficient statistics for precision studies. The reference design adopts a single-phase (SP) readout, where the readout anode is composed of wire planes in the LAr volume. An alternative design is also considered, based on a dual-phase (DP) approach, in which the ionization charges are extracted, amplified and detected in gaseous argon (GAr) above the liquid surface. 
The photon-detection schemes in the two designs are also different, in the SP the photon detectors are distributed within the LAr volume, in the DP they are concentrated at the bottom of the tank. A sketch of both proposals (SP and DP) can be found in Figure \ref{fd-2detectors}.

\begin{figure}
\includegraphics[width=\linewidth]{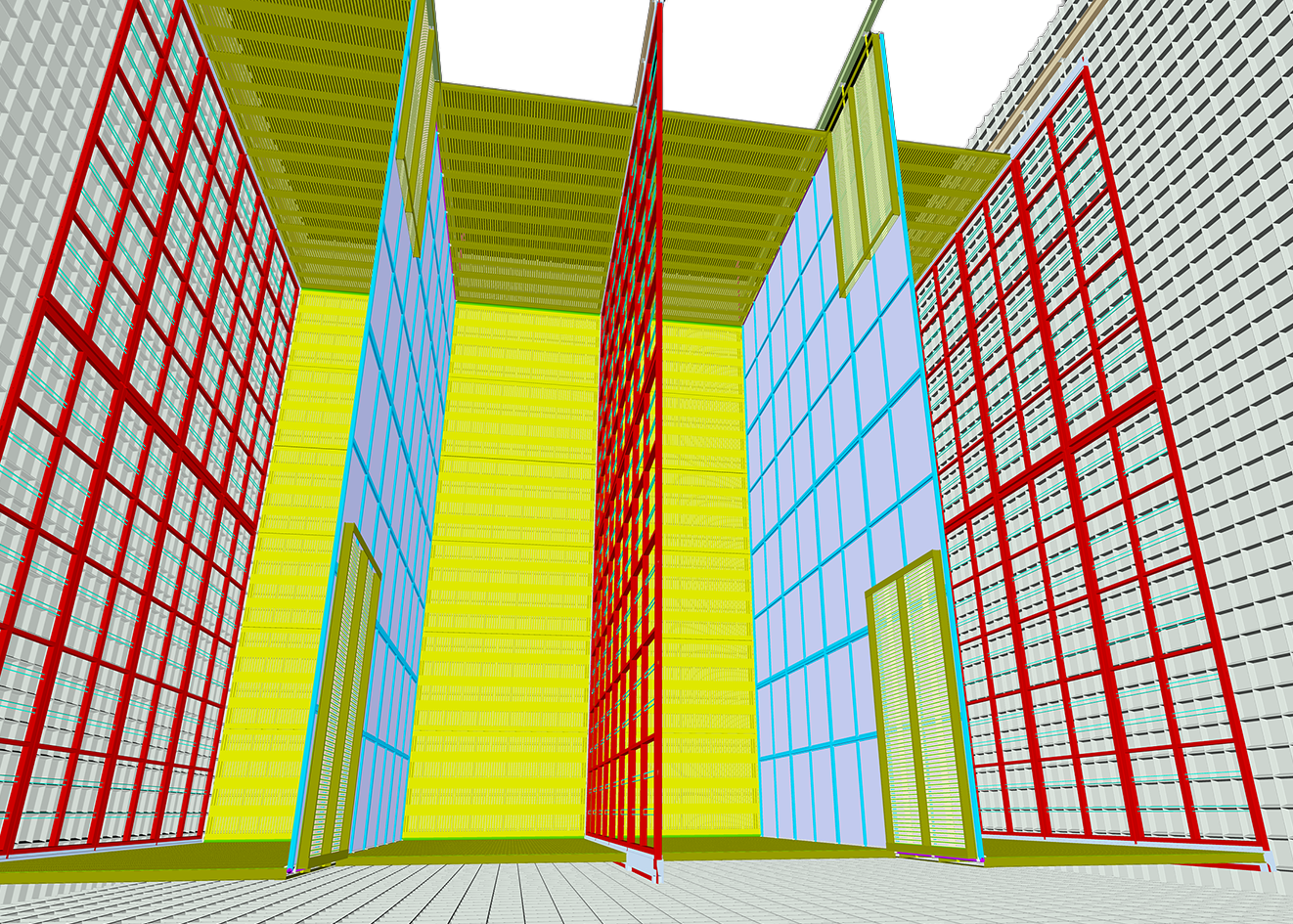}
\caption{Layout of the APAs (blue) and CPAs (red) arrangement inside the LArTPC. The photon detection system will be integrated to the APAs (some frames are shown in green, athe the corners of the APAs).}
\label{tpc-details}
\end{figure}

The 10 kton TPC reference design has an active volume with 12 m high, 14.5 m wide and 58 m long. The TPC is instrumented with anode plane assemblies (APAs), which are 6.3 m high and 2.3 m wide, and cathode plane assemblies (CPAs), 3 m high by 2.3 wide. They are arranged in stacks forming walls (three CPAs interleaved by two APAs) and providing drift modules separated by 3.6 m each along the beam direction (see Figure \ref{tpc-details}). The CPAs are held at −180 kV, such that ionization electrons drift a maximum distance of 3.6 m in the electric field of 500 V/cm. The ultimate validation of the engineered solutions for both designs of the FD is foreseen in the context of the neutrino activities at the CERN around 2018, where full-scale engineering prototypes will be assembled and
commissioned. Following this milestone, a test-beam data campaign will be executed to collect a
large sample of charged-particle interactions in order to study the response of the detector with
high precision. There is recognition that the LArTPC technology will continue to evolve with (1) the large-scale prototypes at the CERN Neutrino
Platform and the experience from the Fermilab SBN program (Acciarri et al. 2015b), and (2) the experience gained during the construction and commissioning of the first 10 kton module. The chosen strategy for implementing the far detector is a staged approach. The deployment of consecutive modules will enable an early science program while allowing implementation of improvements and developments during the experiment’s lifetime. 

\subsection{The Near Detector}

\begin{figure}
\includegraphics[width=\linewidth]{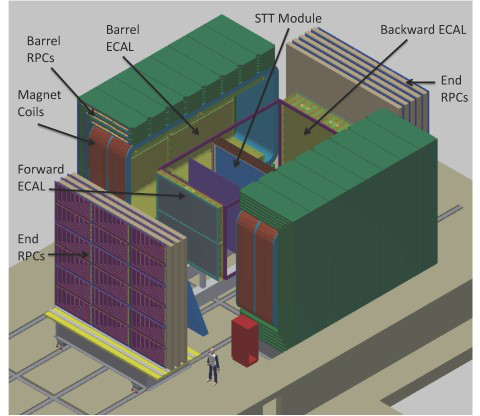}
\caption{ A schematic drawing of the ND fine-grained tracker design.}
\label{ND}
\end{figure}

The primary role of the DUNE near detector system is to perform a precise characterization of the energy spectrum and composition of the neutrino beam at the source, in terms of both muon and electron-flavored neutrinos and antineutrinos. It can also be profited to provide measurements of neutrino interaction cross sections. These features aim to control systematic uncertainties with the precision needed to fulfill the DUNE primary science objectives. The discrimination between fluxes of neutrinos and antineutrinos requires a magnetized neutrino detector to charge-discriminate electrons and muons produced in the neutrino charged-current interactions. As the near detector will be exposed to an intense flux of neutrinos, it will collect an unprecedentedly large sample of neutrino interactions, allowing for an extended
science program. The near detector will, therefore, provide a broad program of fundamental neutrino
interaction measurements, which are an important part of the ancillary scientific goals of the
DUNE collaboration. The reference design for the near detector design is a fine-grained tracker (FGT), illustrated in Figure \ref{ND}. Its subsystems include a central straw-tube
tracker and an electromagnetic calorimeter surrounded by a 0.4 T dipole field. The steel of the magnet yoke will be instrumented with muon identifiers. 

\section{The DUNE physics program}

\subsection{Neutrino Oscillations}

\begin{figure}
\centering
\includegraphics[scale=0.22]{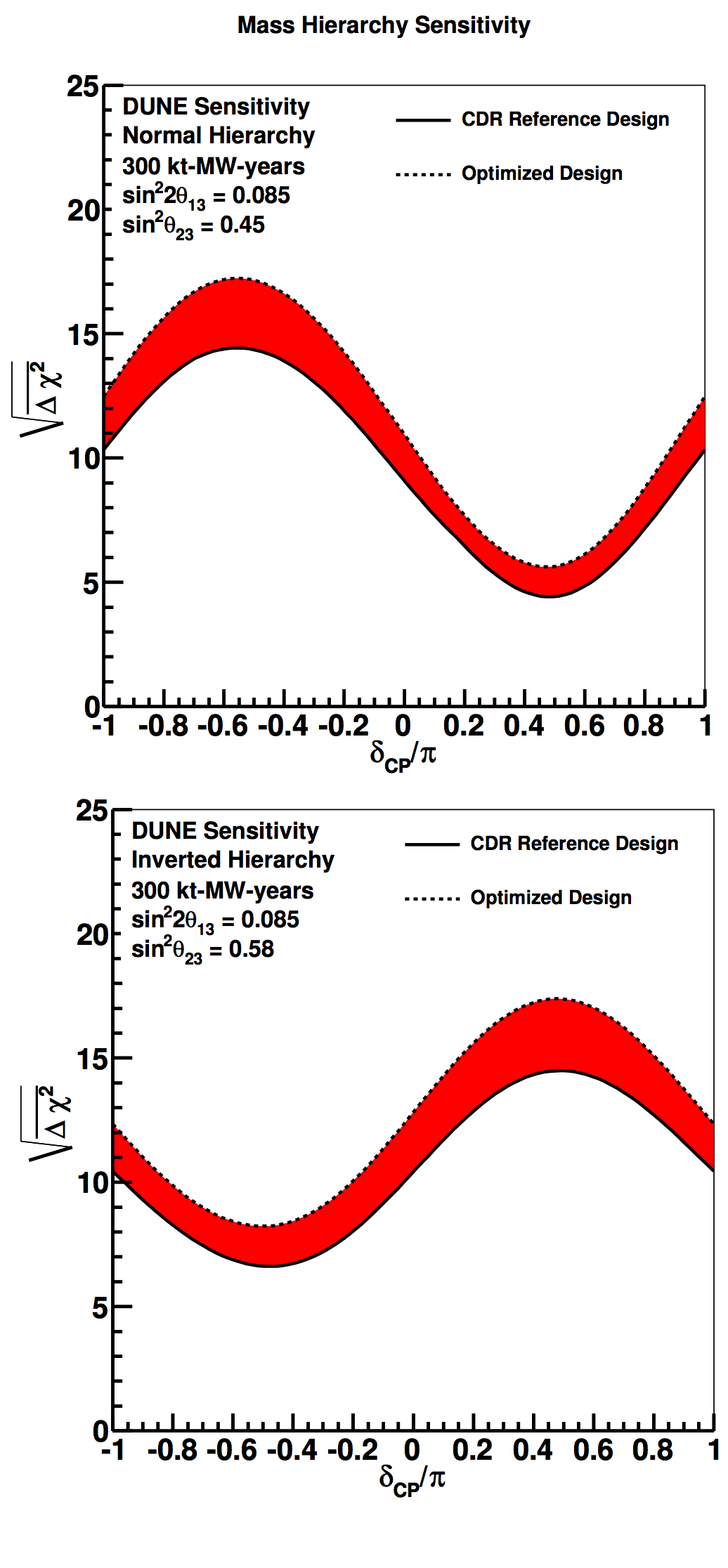}
\caption{The significance with which the mass hierarchy can be determined as a function of the
value of  $\delta_{CP}$ for an exposure of 300 kt $\times$ MW $\times$ year, assuming normal MH (top) or inverted MH (bottom). The shaded region represents the range in sensitivity due to potential variations in the beam design.}
\label{MH}
\end{figure}

The small size of neutrino
masses and their relatively large mixing bears little resemblance to quark masses and mixing,
suggesting that different physics – and possibly different mass scales – in the two sectors may be
present, thus motivating precision study of mixing and CP violation in the lepton sector of the
Standard Model. DUNE plans to pursue a detailed study of neutrino mixing, resolve the neutrino mass ordering, and search for CP violation in the lepton sector by studying the oscillation patterns of high-intensity $\nu_{\mu}$ and $\overline\nu_{\mu}$ beams measured over a long baseline. The oscillation probability of flavor conversion P($\nu_{\mu}\rightarrow \nu_e$), to first order (Nunokawa et al. 2008), considering propagation of a neutrino beam through matter in a constant density approximation, have two major contributions for observations of CP asymmetry: $\delta_{CP}$ and $a=G_f$N$_e / \sqrt{2}$. In \textit{a}, G$_f$ is the Fermi constant, N$_e$ is the is the e$^-$ number density of the matter crossed by neutrinos. 
Both $\delta_{CP}$ and \textit{a} switch signs in going from neutrinos to antineutrinos.
The matter effect is modulated by the value of constant \textit{a}, according to the presence of e$^-$ and the absence of e$^+$. In the few-GeV energy range, the asymmetry from the matter effect
increases with baseline as the neutrinos pass through more matter, therefore an experiment with
a longer baseline will be more sensitive to the neutrino mass hierarchy (MH). For baselines longer than $\sim$1200 km, the degeneracy between the asymmetries from matter and CP-violation effects can be resolved; hence DUNE, with a baseline of ∼1300 km, will be able to unambiguously determine
the neutrino MH and measure the value of $\delta_{CP}$ (Diwan 2004).
The experimental sensitivities presented here are estimated using GLoBES (Huber et al. 2005). GLoBES takes neutrino beam fluxes, cross sections, and detector-response parameterization as inputs. It was also included dependences on the design of the neutrino beam. The cross section inputs to GLoBES have been generated using GENIE 2.8.4 (Andreopoulos 2010). The neutrino oscillation parameters and the uncertainty on those parameters are taken from the Nu-Fit (Gonzalez-Garcia 2014) global fit to neutrino data.
Sensitivities to the neutrino MH and the degree of CP violation are obtained by performing a simultaneous fit over the $\overset{(-)}\nu_{\mu}\rightarrow \overset{(-)}\nu_{\mu}$ and  $\overset{(-)}\nu_{\mu}\rightarrow \overset{(-)}\nu_e$ oscillated spectra. Figure \ref{MH} shows the significance with which the MH can be determined as a function of the value
of $\delta_{CP}$, for an exposure which corresponds to seven years of data (3.5 years in neutrino mode plus 3.5 years in antineutrino mode) with a 40 kton detector and a 1.07 MW (80 GeV) beam. For this exposure, the MH is determined with a minimum significance of $\sqrt{\overline{\Delta\chi^2}}$= 5 for nearly 100\% of $\delta_{CP}$ values for the reference beam design.

\begin{figure}
\centering
\includegraphics[scale=0.22]{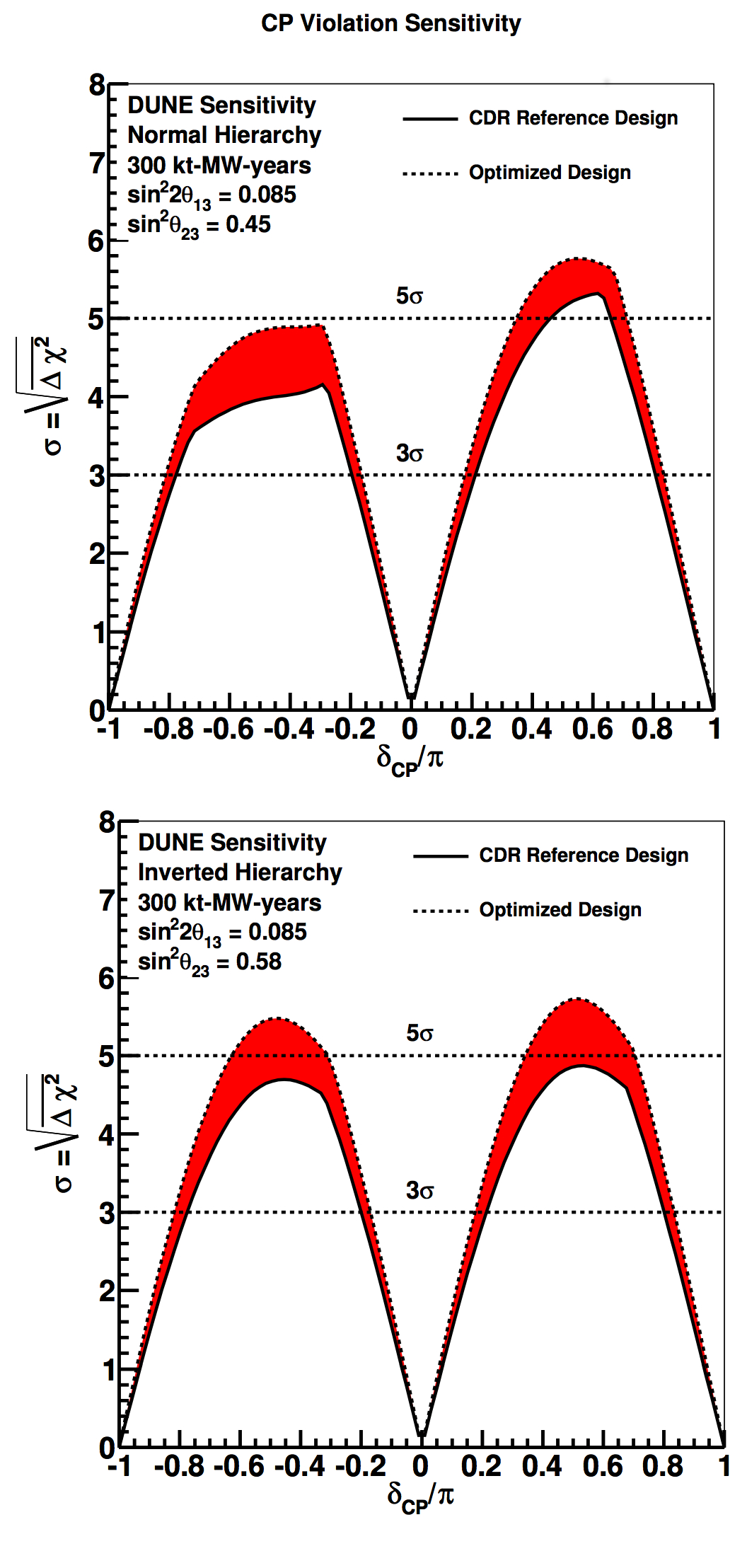}
\caption{The significance with which the CP violation can be determined as a function of the value
of $\delta_{CP}$ for an exposure of 300 kton $\times$ MW $\times$ year assuming normal MH (top) or inverted MH (bottom). The
shaded region represents the range in sensitivity due to potential variations in the beam design.}
\label{deltacp}
\end{figure}

In the approximation for the electron neutrino appearance probability (Nunokawa et al. 2006) there are CP-odd terms (dependent on sin $\delta_{CP}$) that have opposite signs in $\nu_{\mu}\rightarrow \nu_e$ and $\overline\nu_{\mu}\rightarrow \overline\nu_e$ oscillations. For $\delta_{CP} \neq $ 0 or $\pi$, these terms introduce an asymmetry in neutrino versus antineutrino oscillations. 
The variation in the $\overline\nu_{\mu}\rightarrow \overline\nu_e$ oscillation probability (Nunokawa et al. 2006) with the value of $\delta_{CP}$ indicates that it is experimentally possible to measure the value of $\delta_{CP}$ at a fixed baseline using only the observed shape of the $\nu_{\mu}\rightarrow \nu_e$ or the $\overline\nu_{\mu}\rightarrow \overline\nu_e$ appearance signal measured over an energy range
that encompasses at least one full oscillation interval. A measurement of the value of $\delta_{CP} \neq $ 0 or $\pi$,
assuming that neutrino mixing follows the three-flavor model, would imply CP violation. 
Figure \ref{deltacp} shows the significance with which the CP violation ($\delta_{CP} \neq $ 0 or $\pi$) can be determined as a function of the value of $\delta_{CP}$ for an exposure of 300 kt $\times$ MW $\times$ year, which corresponds to seven
years of data (3.5 years in neutrino mode plus 3.5 years in antineutrino mode) with a 40 kton detector
and a 1.07 MW 80 GeV beam.

\subsection{Supernovae Neutrinos}

\begin{figure}
\includegraphics[width=\linewidth]{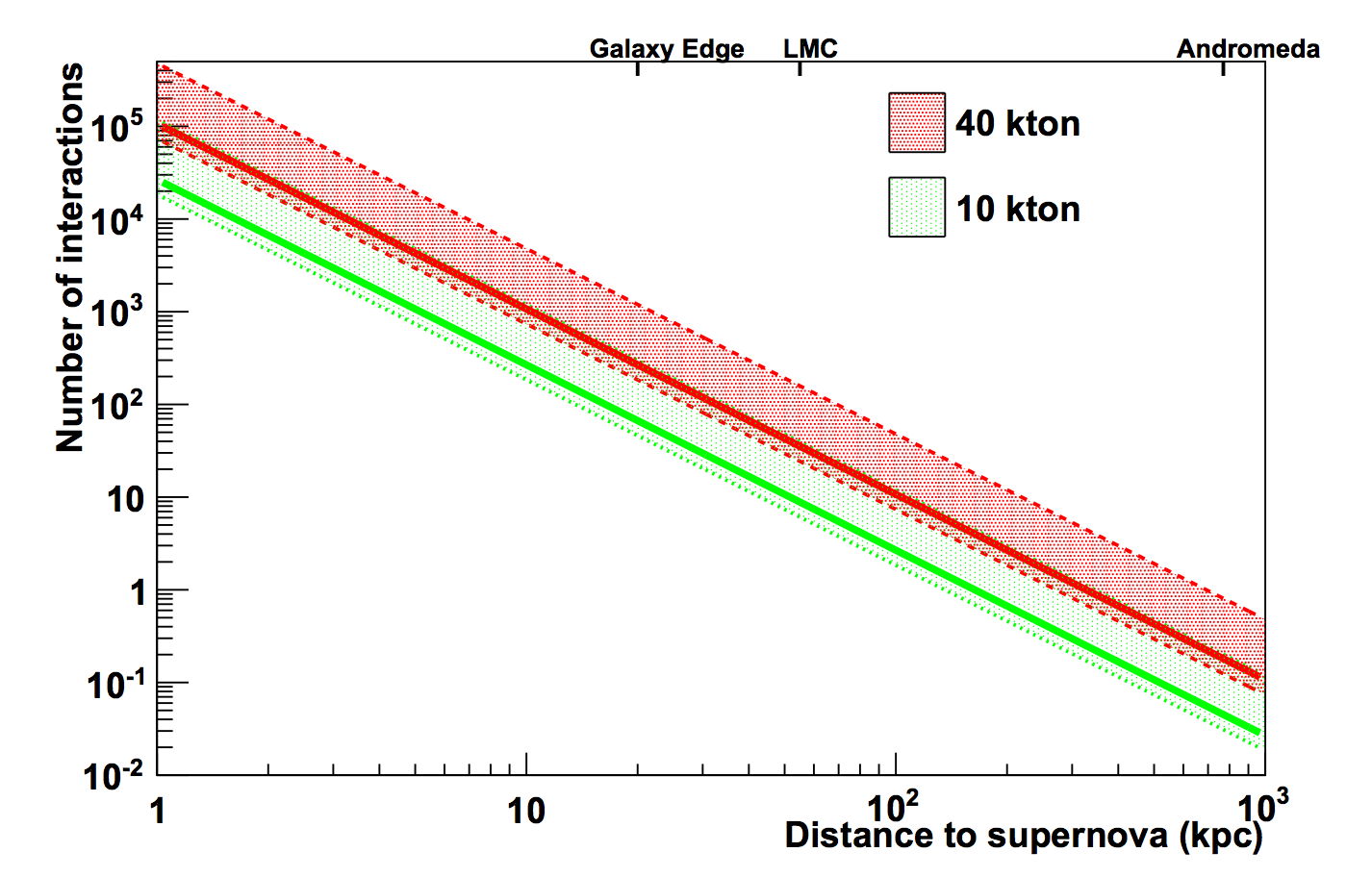}
\caption{ Estimated numbers of supernova neutrino interactions in DUNE ($\nu_e$ events dominate) as a function of distance to
the supernova, for a 10 kton detector (red band) and a 40 kton detector (green band). The width of the bands are the range of possibilities for “Garching-parameterized” (Tamborra 2012)
supernova flux spectra, in both favorable and non-favorable cases. All parameters from neutrino spectrum were considered constant in time. The detector resolution was the same as Icarus and the detection threshold 5 MeV.
}
\label{snrates}
\end{figure}

The DUNE experiment will be sensitive to neutrinos in the few tens of MeV range. This regime is of particular interest for detection of the burst of neutrinos from a galactic core-collapse supernova. The sensitivity of DUNE is primarily to electron flavor supernova neutrinos, and this capability is unique among existing and proposed supernova neutrino detectors for the next decades. Neutrinos from other astrophysical sources are also potentially detectable. 
Liquid argon has a particular sensitivity to the $\nu_e$ component of a supernova neutrino burst, via the charged-current (CC) absorption $\nu_e+^{40}$Ar$\rightarrow$e$^-$+$^{40}$Ar$^*$ for which the observables are the e$^-$ plus de-excitation products from the K$^*$ final state, as well as a $\overline\nu_e$ interaction and elastic scattering on electrons. DUNE’s capability to characterize the $\nu_e$ component of the signal is unique and critical. Other interesting astrophysics studies can be carried out by DUNE, such as solar neutrinos and supernova neutrinos diffuse background, neutrinos from accretion disks and black-hole/neutron star mergers.
There may also be signatures of dark-matter WIMP annihilations in the low-energy signal range.
Neutral-current (NC) scattering on Ar nuclei by any type of neutrino, $\nu_x+^{40}$Ar$\rightarrow\nu_x+^{40}$Ar$^*$, is
another process of interest for supernova detection in LAr detectors although is not yet fully studied. The signature is given by the cascade of de-excitation $\gamma$s from the final-state Ar nucleus that can be potentially be used in tagging NC events.
The predicted event rate (NC or CC) from a supernova burst is calculated by folding expected neutrino differential energy spectra in with cross sections for the relevant channels and with detector response. The number of signal events scales with mass and inverse square of distance as shown in Figure \ref{snrates}. The rates in the Figure \ref{snrates} show the ability of DUNE in have a large statistics in case of a galactic supernova, and resolve astrophysical phenomena to be observable in the neutrino burst signatures. In particular, the supernova explosion mechanism, which in the current paradigm involves energy deposition via neutrinos, is still not well understood, and the neutrinos themselves will bring the insight needed to confirm or refute the paradigm.

\subsection{Baryon Number Violation}

Grand Unified Theories (GUTs) unite the three gauge interactions of particle physics – strong,
weak, and electromagnetic – into one single force. One of the consequences is the predictions about baryon number violation and proton lifetime that may be accessible to DUNE, since the observation requires a kton detector working in low background environment (Senjanovic 2010).  
Although no evidence for proton decay has been detected, lifetime limits from the current generation of experiments constrain the construction of GUT models. In
some cases, these limits are approaching the upper bounds of what these models will allow. This
situation points naturally toward continuing the search with new, highly capable underground detectors, especially those with improved sensitivity to specific proton decay modes favored by GUT models. In particular, the exquisite imaging, particle identification and calorimetric response of the DUNE LArTPC Far Detector opens the possibility of obtaining evidence for nucleon decay on the basis of a single well reconstructed event.
The strength of the DUNE experiment for proton decay studies relies on the capability to detect two significant decay modes: i) p$\rightarrow$e$^+\pi^0$ which is often predicted to have the higher branching fraction. The event kinematics and final states make this channel particularly suitable for a water Cherenkov detectors. ii) p$\rightarrow$K$^+\overline\nu$. This mode is dominant in most supersymmetric GUTs, and is uniquely interesting for DUNE since stopping kaons have a higher ionization density than lower-mass particles. A LArTPC could identify the K$^+$ track with high efficiency. Also, many final states of K$^+$  decay would be fully reconstructible in an LArTPC. The advantage of LArTPC over Cherenkov detectors is clear from the comparison of efficiencies and backgrounds (per kton) for the decays with K in final states. While LArTPC efficiencies are $>$ 95\% in most channels, Cherenkov detectors are in the range of 10\% to 20\%. In LArTPC background are $<$ 2 while in Cherenkov detectors can reach up to 8 counts (Kearns 2013). 

\begin{figure}
\includegraphics[width=\linewidth]{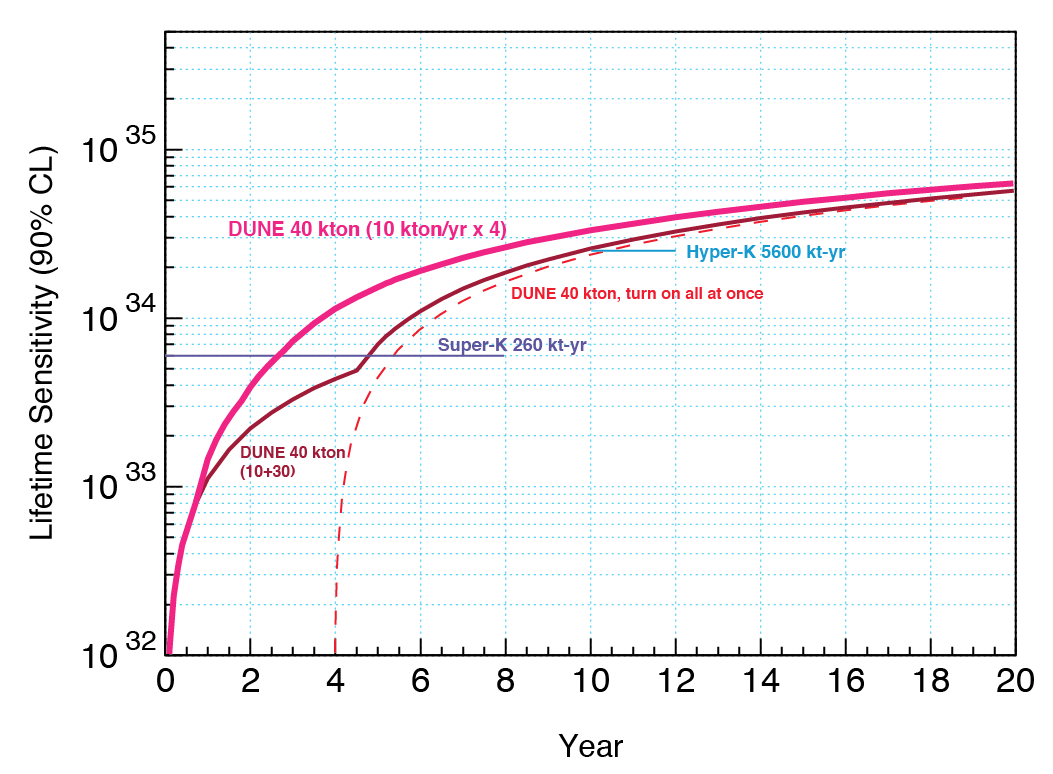}
\caption{ Proton decay lifetime limit for $p\rightarrow$K$^+\overline\nu$ as a function of time for underground LArTPCs, starting with an initial 10~ktons and adding another 10~ktons each year for four years, for a total of 40~ktons. For comparison, the current limit from Super Kamiokande and a projected limit from Hyper Kamiokande are also shown. The limits are at 90\% C.L., calculated for a Poisson process including background, assuming that the detected events equal the expected background.}
\label{pdecay}
\end{figure}

Another promising way of probing baryon number violation in DUNE is through the search
for the spontaneous conversion of neutrons into antineutrons in the nuclear environment. While
these are less well motivated theoretically, opportunistic experimental searches cost little and could have a large payoff. Based on the expected signal efficiency and upper limits on the background rates, the expected limit on the proton lifetime as a function of running time in DUNE for p$\rightarrow$K$^+\overline\nu$ is shown in
Figure \ref{pdecay}.

\section{Timeline}

In the following we summarize some significant milestones for DUNE and LBNF: 

\begin{itemize}

\item[--]
Full-scale prototypes for SP and DP designs working at CERN in 2018.

\item[--]
Installation of the first 10 kton TPC module underground by 2021.

\item[--]
Choosen the technology for the remaining modules (2nd, 3rd and 4th).

\item[--]
Far Detector will start to taking data by 2024 (cosmics).

\item[--]
Far Detector data taking with beam starting at 2026.

\item[--]
Near detector fine grained tracker installed by 2026.

\item[--]
Finish all construction by 2028.

\item[--]
Exposure of 120 kton $\times$ MW $\times$ year by 2035.
\end{itemize}

\section{Conclusions}

The remarkable advances in last decade on the knowledge of neutrino mixing angles and mass splitting paved the way to test the 3-neutrino paradigm, neutrino mass hierarchy and CP asymmetry in the lepton sector.  DUNE will have the key features to successful reach its physics goals: a powerful MW neutrino beam, a highly-capable fine-grained near detector, a massive 40 kton LArTPC working deep underground. In the last years, a strong collaboration has been formed. The strategy for construction has been extensively discussed and provided solid grounds for a clear construction plan.  DUNE and LBNF together consist one of the most ambitious neutrino experiment for the next era of precise measurements. DUNE can shed light on some intriguing opened questions in physics, as the ordering of neutrino mass eigenstates and CP violation in the lepton sector. The expected results are highly significant in statistical terms and can be achieved in reasonable time. Moreover, there is a rich non-oscillation physics program, covering topics as supernovae, nucleon decay, and neutrinos interactions. With DUNE and LBNF we foresee impressive technical and scientific achievements for neutrino physics in the next decades.
\\

\subsection*{\textbf{Acknowledgements}}
  This work was supported by the Brazilians agencies Funda\c{c}\~ao de Amparo a  Pesquisa do Estado de S\~ao Paulo
(FAPESP) and Conselho Nacional de Ci\^encia e Tecnologia (CNPq). 

\newpage

\section*{References}

\def\bibindent{1em}












\end{document}